\documentstyle[prb,preprint,aps]{revtex}

\begin{document}
\title{Electron-phonon scattering in quantum wires exposed to a normal magnetic
field}
\author{Mher M. Aghasyan}
\address{Department of Physics, Yerevan State University, 375049 Yerevan, Armenia}
\author{Samvel M. Badalyan}
\address{Department of Radiophysics, Yerevan State University, \\
375049 Yerevan, Armenia}
\author{Garnett W. Bryant}
\address{National Institute of Standards and Technology, \\
Gaithersburg, MD 20899 USA}
\maketitle

\begin{abstract}
A theory for the relaxation rates of a test electron and electron
temperature in quantum wires due to deformation, piezoelectric acoustical
and polar optical phonon scattering is presented. We represent intra- and
inter-subband relaxation rates as an average of rate kernels weighted by
electron wave functions across a wire. We exploit these expressions to
calculate phonon emission power for electron intra- and inter-subband
transitions in quantum wires formed by a parabolic confining potential. In a
magnetic field free case we have calculated the emission power of acoustical
(deformation and piezoelectric interaction) and polar optical phonons as a
function of the electron initial energy for different values of the
confining potential strength. In quantum wires exposed to the quantizing
magnetic field normal to the wire axis, we have calculated the polar optical
phonon emission power as a function of the electron initial energy and of
the magnetic field.
\end{abstract}

\date{\today}

\newpage

\section{Introduction}

Semiconductor quantum wires attract considerable interest both for
unraveling novel fundamental phenomena and for possible device applications.
Rapid carrier relaxation is crucial for many of technological applications
of these systems, therefore, understanding and characterizing carrier
scattering in quantum wires are important for controlling carrier dynamics
in thermalization, optical, and transport processes.

Theoretically, electron-phonon relaxation in quantum wires has been
addressed in several works \cite
{leburton,fishman,ridley,gold,sarma,shik,constan,mitin,band,peet,pevzner}.
Scattering by optical phonons has been investigated in rectangular wires 
\cite{leburton,sarma}. In cylindrical wires, a simple model with constant
electron wave function inside the wire \cite{fishman}, an infinite and a
finite well confining potential \cite{ridley,constan} have been considered.
Inter-subband scattering has been treated \cite{peet}. Acoustical phonon
relaxation has been studied in wires with parabolic (in one direction) \cite
{shik} and infinite well confining potentials \cite{mitin,band}. Optical
phonon generation has been investigated due to electrophonon resonances\cite
{pevzner}. Although significant progress has been achieved, the problem
still cannot be considered as solved.

In this work we present a theory for calculations of the relaxation rates of
a test electron and electron temperature in quantum wires exposed to a
normal magnetic field. We represent the intra- and inter-subband relaxation
rates as an average of rate kernels weighted by the electron wave functions
across the wire. Exploiting these expressions and the appropriate forms of
the electron subband wave functions, we evaluate the relaxation rates in
quantum wires under different environments. 
We discuss the scattering rates in quantum wires in zero and quantizing
magnetic fields. In the magnetic field free case we present calculations of
the electron scattering rates due to emission of deformation (DA) and
piezoelectric (PA) acoustical, and polar optical (PO) phonons in quantum
wires with a parabolic confining potential as a function of the electron
initial energy for different values of the inter-subband separation. In the
quantizing magnetic field applied normal to the wire axis we study electron
scattering rate due to PO phonon emission as a function of electron initial
energy and of the magnetic field.

\section{Relaxation of a test electron}

In quantum wires, particle motion is described by eigenfunctions $%
|\lambda\rangle\equiv |n l k\rangle=|n l\rangle|k\rangle$ which factor into
subband functions $|nl\rangle=\chi _{nl}({\bf R})$ (${\bf R}=(x,y)$) labeled
by indices $n$ and $l$ corresponding to the lateral quantization across the
wire and into plane waves $|k\rangle=e^{i k z}$ labeled by a wave vector $k$
corresponding to the free translation electron motion along the wire axes $z$%
. The single-particle energy is given by $\varepsilon_{nl}(k)=%
\varepsilon(k)+\varepsilon_{nl}$ where the kinetic energy is $%
\varepsilon(k)=\hbar^2k^2/2m^*$ ($m^*$ is the electron effective mass) and $%
\varepsilon_{nl}$ is the subband energy.

The energy-loss power ${\cal Q}$ for a test electron \cite{gantlev} between
subbands $n,l$ and $n^{\prime },l^{\prime }$ is defined as 
\begin{eqnarray}
{\cal Q}_{n,l\rightarrow n^{\prime },l^{\prime }}^\Upsilon \left(
\varepsilon (k)\right) &=&{\cal Q}_{n,l\rightarrow n^{\prime },l^{\prime
}}^{+\Upsilon }(\varepsilon )-{\cal Q}_{n,l\rightarrow n^{\prime },l^{\prime
}}^{-\Upsilon }(\varepsilon ),  \label{eq1} \\
{\cal Q}_{nl\rightarrow n^{\prime }l^{\prime }}^{\pm \Upsilon }\left(
\varepsilon \right) &=&\sum_{\lambda ^{\prime }}^{(\pm )}\hbar \omega
W_{\lambda \rightarrow \lambda ^{\prime }}^{\pm {\bf q}\Upsilon }{\frac{%
1-f_T\left( \varepsilon \mp \hbar \omega \right) }{1-f_T\left( \varepsilon
\right) }}  \label{eq2}
\end{eqnarray}
where $W_{nlk\rightarrow n^{\prime }l^{\prime }k^{\prime }}^{\pm {\bf q}%
\Upsilon }$ is the scattering probability at which one phonon of the mode $%
\Upsilon $ with the wave vector ${\bf q}=(q_z,{\bf q_{\perp }})$ and the
frequency $\omega =\omega _\Upsilon $ is emitted or absorbed by an electron, 
$f_T$ is the Fermi factor at crystal temperature $T$. The summation $(+)$
and $(-)$ over the final states $\varepsilon (k^{\prime })<\varepsilon (k)$
and $\varepsilon (k^{\prime })>\varepsilon (k)$ corresponds to the phonon
emission and absorption processes, respectively. In the Born approximation
using the explicit form of the transition probability $W_{nlk\rightarrow
n^{\prime }l^{\prime }k^{\prime }}^{\pm {\bf q}\Upsilon }$, we represent the
energy-loss power ${\cal Q}$ in the following general form

\begin{eqnarray}
{\cal Q}_{{n,l\rightarrow n^{\prime },l^{\prime }}}^{\pm \Upsilon }\left(
\varepsilon \right) &=&{\cal Q}_0^\Upsilon \!\int \!\!d^2\!R\!\!\int
\!\!d^2\!R^{\prime }\chi _{n^{\prime }l^{\prime }}^{*}\left( {\bf R^{\prime }%
}\right) \chi _{nl}\left( {\bf R^{\prime }}\right)  \label{eq3} \\
&\times &\chi _{n^{\prime }l^{\prime }}\left( {\bf R}\right) \chi
_{nl}^{*}\left( {\bf R}\right) K_{n,l\rightarrow n^{\prime },l^{\prime
}}^{\pm \Upsilon }\left( \varepsilon (k),\left| {\bf R}-{\bf R}^{\prime
}\right| \right)  \nonumber
\end{eqnarray}
where ${\cal Q}_0^\Upsilon $ is the nominal power and $K_{n,l\rightarrow
n^{\prime },l^{\prime }}^{\pm \Upsilon }$ is a rate kernel which depends on
the type of electron-phonon interaction. By considering different
interaction mechanisms, we obtain the rate kernels $K_{n,l\rightarrow
n^{\prime },l^{\prime }}^{\pm \Upsilon }$ and the nominal powers ${\cal Q}%
_0^\Upsilon $. For $\Upsilon =$ PO phonons: ${\cal Q}_0^{PO}=\hbar \omega
_{PO}/{\overline{\tau }}_{PO}$ and 
\begin{eqnarray}
K_{n,l\rightarrow n^{\prime },l^{\prime }}^{\pm PO}\left( \varepsilon
(k),\left| {\bf R}-{\bf R}^{\prime }\right| \right) &=&\frac{\sqrt{\hbar
\omega _{PO}}}{2\sqrt{\varepsilon +\Delta _{nl,n^{\prime }l^{\prime }}\mp
\hbar \omega _{PO}}}  \label{eq4} \\
&\times &K_0\left( q_z^{\pm }\left| {\bf R}-{\bf R}^{\prime }\right| \right)
\Psi ^{\pm }\left( \varepsilon _{l,n}(k),\hbar \omega _{PO}\right) . 
\nonumber
\end{eqnarray}
For $\Upsilon =$ DA and $\Upsilon =$ PA phonons: ${\cal Q}_0^\Upsilon =\hbar
sp_{_{PO}}/{\overline{\tau }}_\Upsilon $ and 
\begin{eqnarray}
K_{n,l\rightarrow n^{\prime },l^{\prime }}^{\pm \Upsilon }\left( \varepsilon
,\left| {\bf R}-{\bf R}^{\prime }\right| \right) &=&\frac{\sqrt{2ms^2}}{%
2(sp_{_{PO}})^{2+\sigma _\Upsilon }}  \label{eq5} \\
&\times &\int\limits_{\omega _1^{\pm }}^{\omega _2^{\pm }}\frac{d\omega
\omega ^{1+\sigma _\Upsilon }J_0\left( q_{\perp }^{\pm }\left| {\bf R}-{\bf R%
}^{\prime }\right| \right) }{\sqrt{\varepsilon +\Delta _{_{ln,l^{\prime
}n^{\prime }}}\mp \hbar \omega }}\Psi ^{\pm }\left( \varepsilon
_{ln}(k);\hbar \omega \right) ,  \nonumber
\end{eqnarray}
with $\sigma _{_{DA}}=2$ and $\sigma _{_{PA}}=0$. Here the nominal
scattering times are given by \cite{gantlev} 
\begin{equation}
\frac 1{\overline{\tau }_{_{PO}}}=2\alpha _{_{PO}}\omega _{_{PO}},\frac 1{%
\overline{\tau }_{_{DA}}}=\frac{\Xi ^2p_{_{PO}}^3}{2\pi \hbar \varrho s^2},%
\frac 1{\overline{\tau }_{_{PA}}}=\frac{(e\beta )^2p_{_{PO}}}{2\pi \hbar
\varrho s^2}  \label{eq6}
\end{equation}
where $\alpha _{_{PO}}$ is the Fr\"{o}hlich coupling, $\Xi $ and $e\beta $
are the deformation and piezoelectric potential constants, $\varrho _0$ is
the crystal mass density, $s$ the sound velocity, $\hbar p_{_{PO}}=\sqrt{%
2m\hbar \omega _{_{PO}}}$, and $\omega _{_{PO}}$ the polar optical phonon
frequency. In Eqs.\ (\ref{eq4}) and (\ref{eq5}) $\Delta _{nl,n^{\prime
}l^{\prime }}=\varepsilon _{l,n}-\varepsilon _{l^{\prime },n^{\prime }}$, $%
J_0$ is the Bessel function of the first kind and $K_0$ the modified Bessel
function of the second kind, the function $\Psi ^{\pm }$ is 
\begin{equation}
\Psi ^{\pm }\left( x,y\right) =\left( N_T(y)+\frac 12\pm \frac 12\right) 
\frac{1-f_T\left( x\mp y\right) }{1-f_T\left( x\right) }  \label{eq7}
\end{equation}
where $N_T$ is the Bose factor. The phonon momenta are given by 
\begin{eqnarray}
q_z^{\pm } &=&\sqrt{\frac{2m}{\hbar ^2}}\left| \sqrt{\varepsilon (k)}-r\sqrt{%
\varepsilon (k)+\Delta _{nl,n^{\prime }l^{\prime }}\mp \hbar \omega _{_{PO}}}%
\right| ,  \label{eq8} \\
q_{\perp }^{\pm } &=&\left| \frac{\omega ^2}{s^2}-\left( q_z^{\pm }\right)
^2\right| ^{1/2}.  \label{eq9}
\end{eqnarray}
In Eq.\ (\ref{eq9}), $q_z$ should be taken from Eq.\ (\ref{eq8}) with $%
\omega _{_{PO}}$ replaced by $\omega $, $r=+1$ ( $r=-1$) corresponds to
scattering (backscattering) processes when the electron momentum $k$ does
not (does) change its direction. The limits of integration in Eq.\ (\ref{eq5}%
) are $\omega _2^{-}=\infty ,\omega _2^{+}=\varepsilon (k_z)+\Delta
_{nl,n^{\prime }l^{\prime }}/{\hbar }$ . $\omega _1^{^{\pm }}$ are solutions
of the equation $q_{\perp }^{\pm }=0$. For intra-subband scattering $\omega
_1^{^{\pm }}=0$ while inter-subband scattering always has a threshold, $%
\omega _1^{^{\pm }}\neq 0$. The existence of this threshold is important
especially at low temperatures when typical electron energies in scattering
are $\varepsilon (k)\sim \omega _1^{^{\pm }}$. If either $\varepsilon (k)$
or $\Delta _{nl,n^{\prime }l^{\prime }}$ are much larger than $\omega
_1^{^{\pm }}$ then one can find the following analytical approximation 
\begin{equation}
\omega _1^{^{\pm }}\approx {\frac 1\hbar }{\frac{\left| \sqrt{\varepsilon (k)%
}-r\sqrt{\varepsilon (k)+\Delta _{nl,n^{\prime }l^{\prime }}}\right| }{%
\left( \sqrt{2ms^2}\right) ^{-1}\mp r\left( 2\sqrt{\varepsilon (k)+\Delta
_{nl,n^{\prime }l^{\prime }}}\right) ^{-1}}.}  \label{eq10}
\end{equation}
Thus, Eqs. (\ref{eq1})-(\ref{eq3}) with the rate kernels given by Eqs.\ (\ref
{eq4}) and (\ref{eq5}) provide a new approach to calculate the energy-loss
power due to PO, PA, and DA phonon scattering in quantum wires with an
arbitrary cross section and under different environments.

\section{Electron temperature relaxation}

If the distribution of hot electrons can be described by an electron
temperature $T_e>T$, we can determine the energy relaxation rate for the
whole electron gas. In this case electron temperature relaxation between
subbands $n,l$ and $n^{\prime },l^{\prime }$ can be described by the
energy-loss power per electron \cite{gantlev} which is given by 
\begin{eqnarray}
\overline{Q}_{n,l\rightarrow n^{\prime },l^{\prime }}^\Upsilon (T_e,T) &=&%
\overline{Q}_{_{n,l\rightarrow n^{\prime },l^{\prime }}}^{+\Upsilon }-%
\overline{Q}_{_{n,l\rightarrow n^{\prime },l^{\prime }}}^{-\Upsilon },
\label{eq11} \\
\overline{{\cal Q}}_{n,l\rightarrow n^{\prime },l^{\prime }}^\Upsilon
(T_e,T) &=&{\frac 1{N_1L}}\sum_{{n,l,k}}\,f_{T_e}(\varepsilon )\sum_{{%
n^{\prime },l^{\prime }k^{\prime }}}^{(\pm )}\,\hbar \omega \,  \label{eq12}
\\
&\times &W_{nlk\rightarrow n^{\prime }l^{\prime }k^{\prime }}^{\pm {\bf q}%
\Upsilon }[1-f_{T_e}(\varepsilon \mp \hbar \omega )].  \nonumber
\end{eqnarray}
Here $N_1$ is the electron linear concentration. Direct calculations show
that to obtain $\overline{Q}_{_{n,l\rightarrow n^{\prime },l^{\prime
}}}^{\pm \Upsilon }$ one can use Eq.\ (\ref{eq3}) but with the kernel $%
K_{n,l\rightarrow n^{\prime },l^{\prime }}^{\pm \Upsilon }$ replaced by the
average rate kernel $\overline{K}_{n,l\rightarrow n^{\prime },l^{\prime
}}^{\pm \Upsilon }$. For $\Upsilon =$ PO phonons, we obtain 
\begin{eqnarray}
\overline{K}_{n,l\rightarrow n^{\prime },l^{\prime }}^{\pm PO}\left(
T_e,T;\left| {\bf R}-{\bf R}^{\prime }\right| \right)  &=&\frac{p_{_{PO}}}{%
2\pi N_1}\int\limits_0^\infty \frac{d\varepsilon (k)}{\sqrt{\varepsilon (k)}}
\label{eq13} \\
&\times &\frac{K_0\left( q_z^{\pm }\left| {\bf R}-{\bf R}^{\prime }\right|
\right) \Phi ^{\pm }\left( \varepsilon _{l,n}(k),\hbar \omega _{PO}\right) }{%
\sqrt{\varepsilon (k)+\Delta _{_{ln,l^{\prime }n^{\prime }}}\mp \hbar \omega
_{PO}}}.  \nonumber
\end{eqnarray}
For $\Upsilon =$ DA and $\Upsilon =$ PA phonons 
\begin{eqnarray}
\overline{K}_{n,l\rightarrow n^{\prime },l^{\prime }}^{\pm \Upsilon }\left(
T_e,T;\left| {\bf R}-{\bf R}^{\prime }\right| \right)  &=&\frac{ms}{2\pi
\hbar N_1(sp_{_{PO}})^{2+\sigma _\Upsilon }}\int\limits_0^\infty \frac{%
d\varepsilon (k)}{\sqrt{\varepsilon (k)}}\int\limits_{\omega _1^{\pm
}}^{\omega _2^{\pm }}\frac{\omega ^3d\omega }{\sqrt{\varepsilon (k)+\Delta
_{_{ln,l^{\prime }n^{\prime }}}\mp \hbar \omega }}  \label{eq14} \\
&\times &J_0\left( q_{\perp }^{\pm }\left| {\bf R}-{\bf R}^{\prime }\right|
\right) \Phi ^{\pm }(\varepsilon _{l,n}(k);\hbar \omega )  \nonumber
\end{eqnarray}
where $\Phi $ is the following function

\begin{eqnarray}
\Phi ^{\pm }(x;y) &=&\left( N_T(y)+\frac 12\pm \frac 12\right)   \label{eq15}
\\
&\times &f_{T_e}\left( x\right) \left( 1-f_{T_e}\left( x\mp y\right) \right) 
\nonumber
\end{eqnarray}

\section{Scattering in the zero magnetic field}

In quantum wires with confining potential of cylindrical symmetry, the
electron wave functions in the absence of the magnetic field in the plane
perpendicular to the wire axis are represented in the form. 
\begin{equation}
\chi _{nl}\left( {\bf R}\right) =\frac 1{\sqrt{2\pi }}e^{il\varphi }\chi
_{nl}\left( R\right) .  \label{eq16}
\end{equation}
Substituting this into Eq.\ (\ref{eq3}) and integrating over $\varphi $, we
represent the energy-loss power in the form 
\begin{eqnarray}
{\cal Q}_{{n,l\rightarrow n^{\prime },l^{\prime }}}^{\pm \Upsilon }\left(
\varepsilon \right) &=&{\cal Q}_0^\Upsilon \!\int \!RdR\!\!\int \!R^{\prime
}dR^{\prime }\chi _{n^{\prime }l^{\prime }}^{*}\left( R^{\prime }\right)
\chi _{nl}\left( R^{\prime }\right)  \label{eq17} \\
&\times &\chi _{n^{\prime }l^{\prime }}\left( R\right) \chi _{nl}^{*}\left(
R\right) \left( K_{n,l\rightarrow n^{\prime },l^{\prime }}^{\pm \Upsilon
}\left( \varepsilon (k);R,{R}^{\prime }\right) \right) _{cyl}  \nonumber
\end{eqnarray}
where the rate kernel factors into a product of two functions, each of them
depends only on the modulus $R$ or $R^{\prime }$. We find for PO interaction
that the rate kernels $\left( K_{n,l\rightarrow n^{\prime },l^{\prime
}}^{\pm PO}\right) _{cyl}$ and $\left( \overline{K}_{n,l\rightarrow
n^{\prime },l^{\prime }}^{\pm PO}\right) _{cyl}$ for a test electron and for
electron temperature relaxation can be obtained by the replacement 
\[
K_0\left( q_z^{\pm }\left| {\bf R}-{\bf R}^{\prime }\right| \right) \to
K_{l-l^{\prime }}\left( q_z^{\pm }R\right) I_{l-l^{\prime }}\left( q_z^{\pm
}R^{\prime }\right) 
\]
in Eqs.\ (\ref{eq4}) and (\ref{eq13}), respectively. Here $I_l$ is the
modified Bessel function of the first kind and $R>R^{\prime }$. To obtain
the kernels $\left( K_{n,l\rightarrow n^{\prime },l^{\prime }}^{\pm
PA,DA}\right) _{cyl}$ and $\left( \overline{K}_{n,l\rightarrow n^{\prime
},l^{\prime }}^{\pm PA,DA}\right) _{cyl}$, the replacement 
\[
J_0\left( q_z^{\pm }\left| {\bf R}-{\bf R}^{\prime }\right| \right) \to
J_{l-l^{\prime }}\left( q_z^{\pm }R\right) J_{l-l^{\prime }}\left( q_z^{\pm
}R^{\prime }\right) 
\]
should be done in Eqs.\ (\ref{eq5}) and (\ref{eq14}), respectively.

For the parabolic confining potential $V(R)=m^{*}\omega _0^2R^2/2$, the
electron wave functions are given by \cite{fock} 
\[
\chi _{nl}\left( R\right) =\sqrt{\frac{2n!}{(n+|l|)!}}\frac 1{a_0}e^{-\frac{%
r^2}{2a_0^2}}\left( \frac r{a_0}\right) ^{|l|}L_n^{|l|}\left( \frac{r^2}{%
a_0^2}\right) . 
\]
The subband energy $\varepsilon _{n,l}=\left( 2n+|l|+1\right) \hbar \omega
_0 $ where $\omega _0$ is confining potential strength, $a_0=\sqrt{\hbar
/(m^{*}\omega _0)}$, $L_n^{|l|}(x)$ gives the generalized Laguerre
polynomial. Using these functions we have calculated PO, PA, and DA emission
power for electron transitions between subbands $n,l$ and $n^{\prime
},l^{\prime }$ with $n,l=0,1$ as a function of the electron initial energy
for different values of $\omega _0$. In Figs.\ \ref{fg1} and \ref{fg2} we
present the results of calculations. It is seen from Fig.\ \ref{fg1} that
the PO phonon emission rate diverges at $\varepsilon =\Delta _{nl,n^{\prime
}l^{\prime }}+\hbar \omega _{PO}$ due to transitions with the electron final
states at the subband bottom where the 1D density of states has a
square-root singularity. For energies far from $\Delta _{nl,n^{\prime
}l^{\prime }}+\hbar \omega _{PO}$ we find a weak dependence of the emission
rate on $\varepsilon (k)$ while there is a strong dependence on the subband
separation $\omega _0$ and the quantum numbers $n,l$. The intra-subband
acoustical phonon emission rate has a peak at small energies (Fig.\ \ref{fg2}%
) while inter-subband emission is finite even at $\varepsilon (k)=0$. The
peak position decreases with an increase of $a_0$. We find that
intra-subband DA phonon emission is weaker than PA phonon emission. This
difference is less pronounced at inter-subband emission.

\section{Scattering in the magnetic field normal to the wire axis}

When the magnetic field is applied perpendicular to the wire axis, the
electron energy and wave functions in the parabolic confining potential $%
V(x)=m\omega _x^2x^2/2$ and $V(y)=m\omega _y^2y^2/2$ are given by\cite
{childers}

\begin{equation}
\chi _{n,l}({\bf R})=\frac{\exp \left[ -\frac{(x-x_0)^2}{2a_x^2}\right]
H_n\left( \frac{x-x_0}{a_x}\right) }{\sqrt{2^nn!a_x\sqrt{\pi }}}\frac{\exp
\left[ -\frac{y^2}{2a_y^2}\right] H_l\left( \frac y{a_y}\right) }{\sqrt{%
2^ll!a_y\sqrt{\pi }}},  \label{eq18}
\end{equation}

\begin{equation}
\varepsilon _{n,l}(k)=\varepsilon _B(k)+\hbar \Omega _x\left( n+\frac 12%
\right) +\hbar \omega _y\left( l+\frac 12\right) ,\varepsilon _B(k)=\frac{%
\hbar ^2k^2}{2m_B}  \label{eq19}
\end{equation}
where $a_x=\sqrt{\hbar /m^{*}\Omega _x}$, $a_y=\sqrt{\hbar /m^{*}\omega _y}$%
, $\Omega _x^2=\omega _x^2+\omega _B^2$, $m_B=m^{*}\Omega _x^2/\omega _x^2$, 
$\omega _B=eB/m^{*}c$, and $H_n$ gives the Hermite polynomials.

In this case to obtain the rate kernels $K^{\pm \Upsilon }$ we should
multiply Eqs.\ (\ref{eq4}), (\ref{eq5}), (\ref{eq13}), and (\ref{eq14}) by a
factor $\Omega _x/\omega _x$ and replace $\varepsilon \left( k\right) $ by $%
\varepsilon _B\left( k\right) $ in these equations. Below we will discuss
only electron PO phonon scattering. Scattering by acoustical phonons in the
normal magnetic field has been studied by Shik and Challis \cite{shik}.

Substituting the kernel $K^{PO}$ and wave functions (\ref{eq18}) in Eq. (\ref
{eq3}), we represent the PO phonon emission power in the form

\begin{equation}
Q_{n,l\rightarrow n^{\prime },l^{\prime }}^{PO}=Q_0^{PO}\frac{\sqrt{\hbar
\omega _{PO}}}{\sqrt{\varepsilon _B+\Delta _{n,l,n^{\prime },l^{\prime
}}-\hbar \omega _{PO}}}\frac{\Omega _x}{\omega _x}I_{nl}^{n^{\prime
}l^{\prime }}  \label{eq20}
\end{equation}
where the form factors $I_{nl}^{n^{\prime }l^{\prime }}$ for the most
important intra-subband $00\rightarrow 00$ and inter-subband $10\rightarrow
00$ transitions are reduced to the following one-dimensional integrals

\begin{equation}
I_{00}^{00}=\int_0^\infty \frac{e^{-\zeta }d\zeta }{\left( 2\zeta
+q_z^2a_x^2\right) \left( 2\zeta +q_z^2a_y^2\right) },  \label{eq21}
\end{equation}

\begin{equation}
I_{10}^{00}=I_{00}^{00}+\frac 2{q_z^2a_x^2-q_z^2a_y^2}\int_0^\infty \left(
1-\zeta \right) e^{-\zeta }\sqrt{\frac{2\zeta +q_z^2a_y^2}{2\zeta +q_z^2a_x^2%
}}d\zeta  \label{eq22}
\end{equation}
which we calculate numerically. The results of calculation are shown in
Figs. \ref{fg3}-\ref{fg7}. The diagrams of Figs. \ref{fg3} and \ref{fg4}
represent the intra-subband PO phonon emission power dependencies on the
electron initial energy and on the magnetic field, respectively, for several
values of the confining potential strengths $\omega _x$ and $\omega _y$. It
is seen from both figures that, as it was in the magnetic field free case,
the PO phonon emission power diverges at the phonon emission threshold which
is given in this case by $\varepsilon _B\left( k\right) =\hbar \omega _{PO}$%
. Because of the electron mass dependence on the magnetic field, the
threshold values are trivially shifted to higher electron energies with an
increase of the magnetic field. At energies far from the threshold, the
effect of the magnetic field is weak. At inter-subband transitions, $\Delta
_{10,00}$ differs from zero and depends on the magnetic field. In this case
there is a threshold line given by $\varepsilon _B\left( k\right) =\hbar
\omega _{PO}-\Delta _{10,00}$ and shown in Fig. \ref{fg5}. The electron
threshold energy increases in the magnetic field from $\varepsilon _0=\hbar
(\omega _{PO}-\omega _x)$ at $B=0$ up to the value $\varepsilon _1$ at $%
B=B_1 $ ($B_1$ is determined from threshold conditions $\Omega _x=2\omega
_{PO}/3$). For magnetic fields larger than $B_1$ the threshold energy
decreases and vanishes at $B=B_{PO}$ ($B_{PO}$ is determined from the
resonance $\Omega _x=\omega _{PO}$). According to this the emission power
dependence on the initial energy has no divergence for $B=21,22$ and $25$ T
(see Fig. \ref{fg6}). For magnetic fields larger but near $B_{PO}$, $Q^{PO}$
has a peak for small values of $\varepsilon $ while for magnetic fields far
from $B_{PO}$, $Q^{PO}$ increases monotonically in $\varepsilon $. For a
given value of $\varepsilon <\varepsilon _1$, there is an interval of
magnetic fields where PO phonon emission is not possible (Fig. \ref{fg5})
while $Q^{PO}$diverges at the edges of this interval (Fig. \ref{fg7}). This
interval vanishes at $\varepsilon =\varepsilon _1$ so that at energies
larger but not far from $\varepsilon _1$, $Q^{PO}$ as a function of the
magnetic field has a peak at $B=B_1$ (Fig. \ref{fg7}). The second peak in
the magnetic field dependence of the PO phonon emission power occurs at the
resonance field $B_{PO}$ and corresponds to the vertical electron
transitions with the phonon momentum $q_z=0$.

In conclusion, we have presented a theory for the test carrier and carrier
temperature relaxation rates in quantum wires. This theory has been
exploited to calculate the PO, PA and DA phonon emission power for electron
intra- and inter-subband transitions in quantum wires with the parabolic
confining potential for different values of the potential strength. We have
discussed the phonon emission power in quantum wires in the zero and
quantizing magnetic field normal to the wire axis.\bigskip

\acknowledgements
S.M.B. and M.M.A. would like to thank A. A. Kirakossyan for for useful
discussion. The research described in this publication was made possible in
part by Award No. 375100 of the U. S. Civilian Research \& Development
Foundation for Independent States of the Former Soviet Union (CRDF). S.M.B.
acknowledge partial support from the Belgian Federal Office for Scientific,
Technical, and Cultural Affairs.

\begin{figure}[tbp]
\caption{The PO phonon emission power versus the electron initial kinetic
energy in zero magnetic field at the intra- and inter-subband electron
transitions and for different values of the subband separation $%
\hbar\omega_0 $.}
\label{fg1}
\end{figure}

\begin{figure}[tbp]
\caption{The PA and DA phonon emission power versus the electron initial
kinetic energy in zero magnetic field at the intra- and inter-subband
electron transitions and for different values of the subband separation $%
\hbar\omega_0$.}
\label{fg2}
\end{figure}

\begin{figure}[tbp]
\caption{The PO phonon emission power dependence on the electron initial
kinetic energy in the quantizing magnetic field at the intra-subband
electron transitions for various values of the subband separations $%
\hbar\omega_x$ and $\hbar\omega_y$.}
\label{fg3}
\end{figure}

\begin{figure}[tbp]
\caption{The PO phonon emission power dependence on the magnetic field at
the intra-subband electron transitions for various values of the subband
separations $\hbar\omega_x$ and $\hbar\omega_y$.}
\label{fg4}
\end{figure}

\begin{figure}[tbp]
\caption{The threshold line in the ($\varepsilon,B$)-plane which separates
regions where PO phonon emission is and is not possible. $%
\varepsilon_0/\hbar \omega_{PO}=0.84$, $\varepsilon_1/\hbar \omega_{PO}=5.97$%
, $B_1=13.64$ T, $B_{PO}=20.67$ T.}
\label{fg5}
\end{figure}

\begin{figure}[tbp]
\caption{The PO phonon emission power dependence on the electron initial
kinetic energy at the inter-subband electron transitions for various values
of the magnetic field.}
\label{fg6}
\end{figure}
\begin{figure}[tbp]
\caption{The PO phonon emission power dependence on the magnetic field at
the inter-subband electron transitions for various values of electron
initial kinetic energy.}
\label{fg7}
\end{figure}

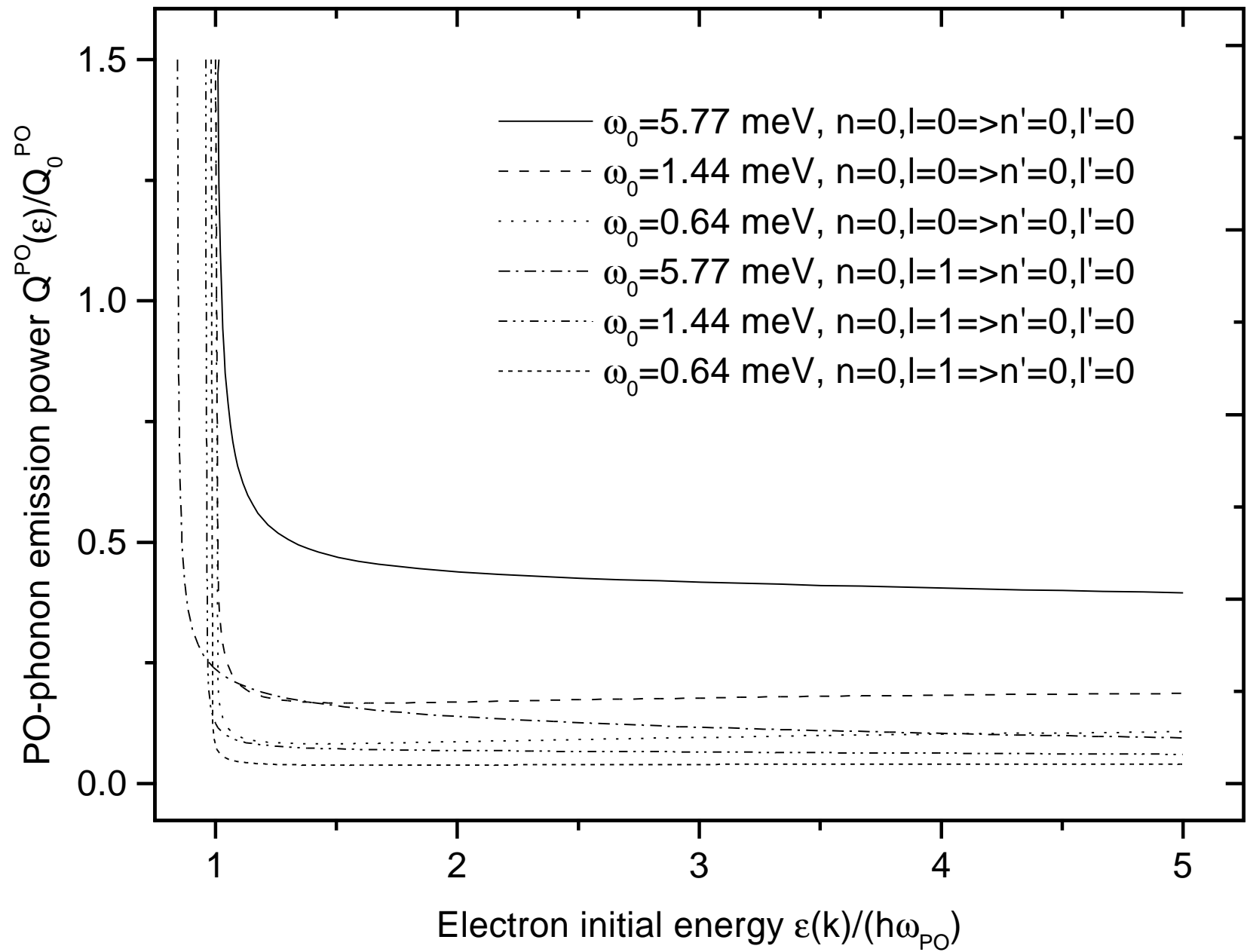
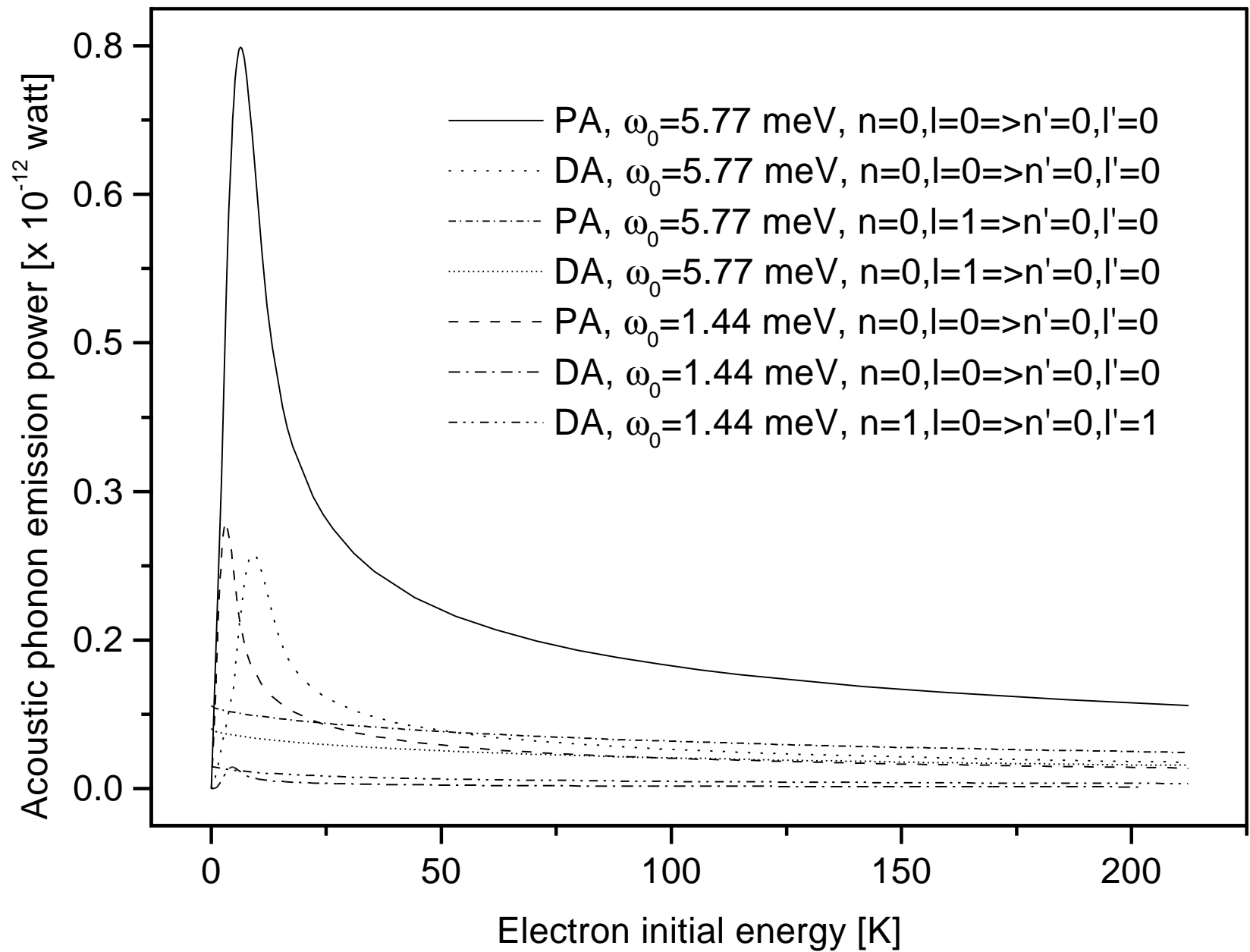
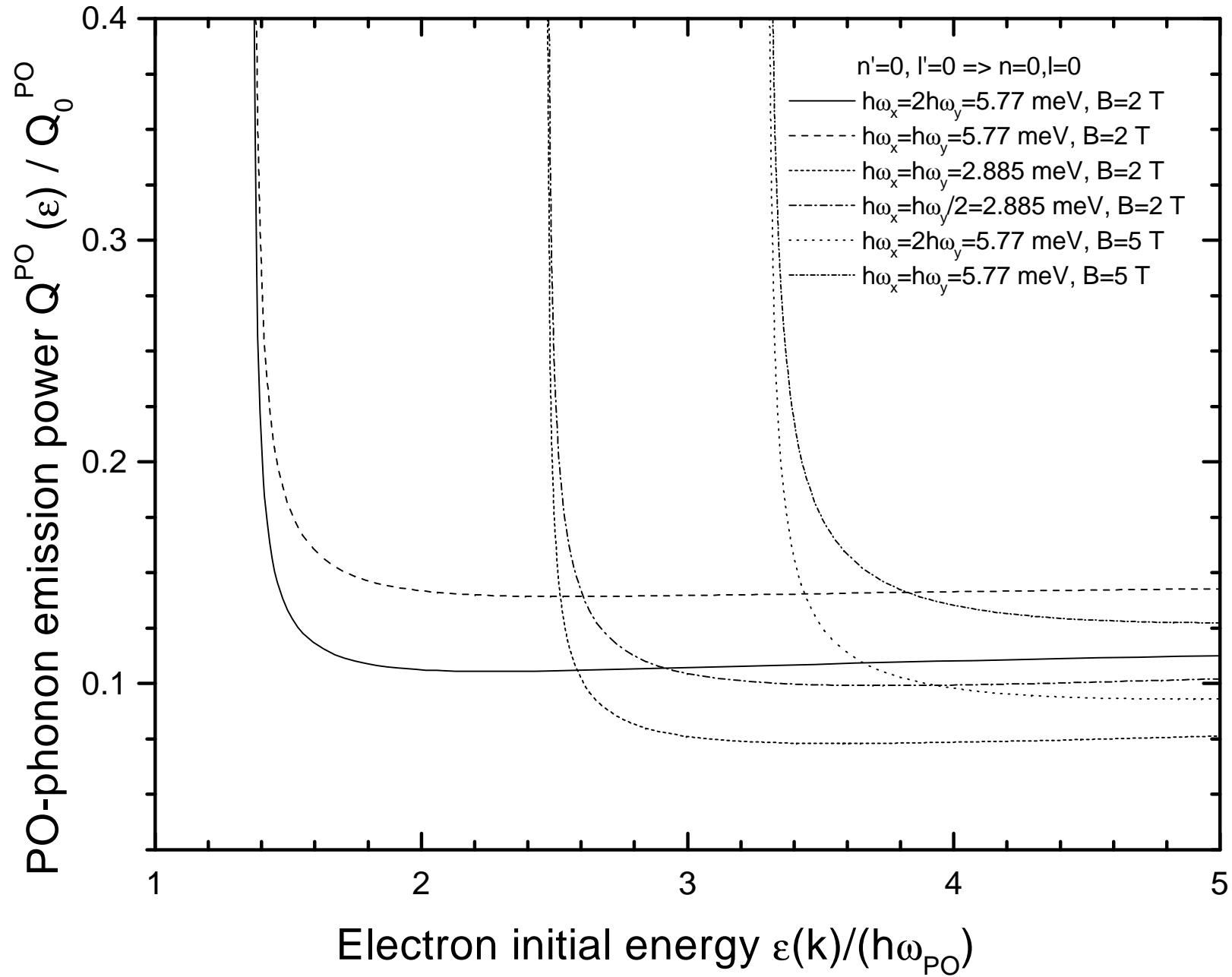
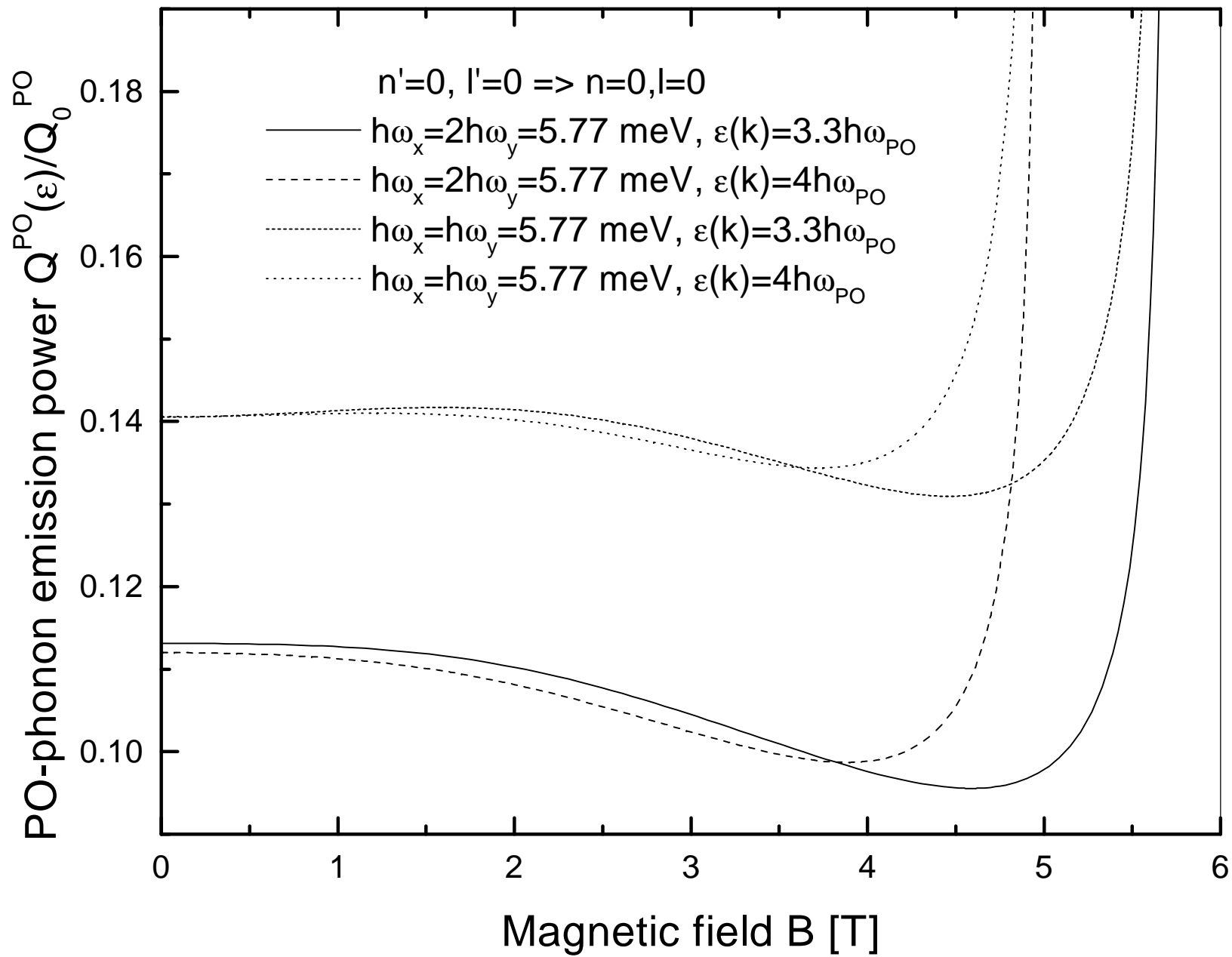
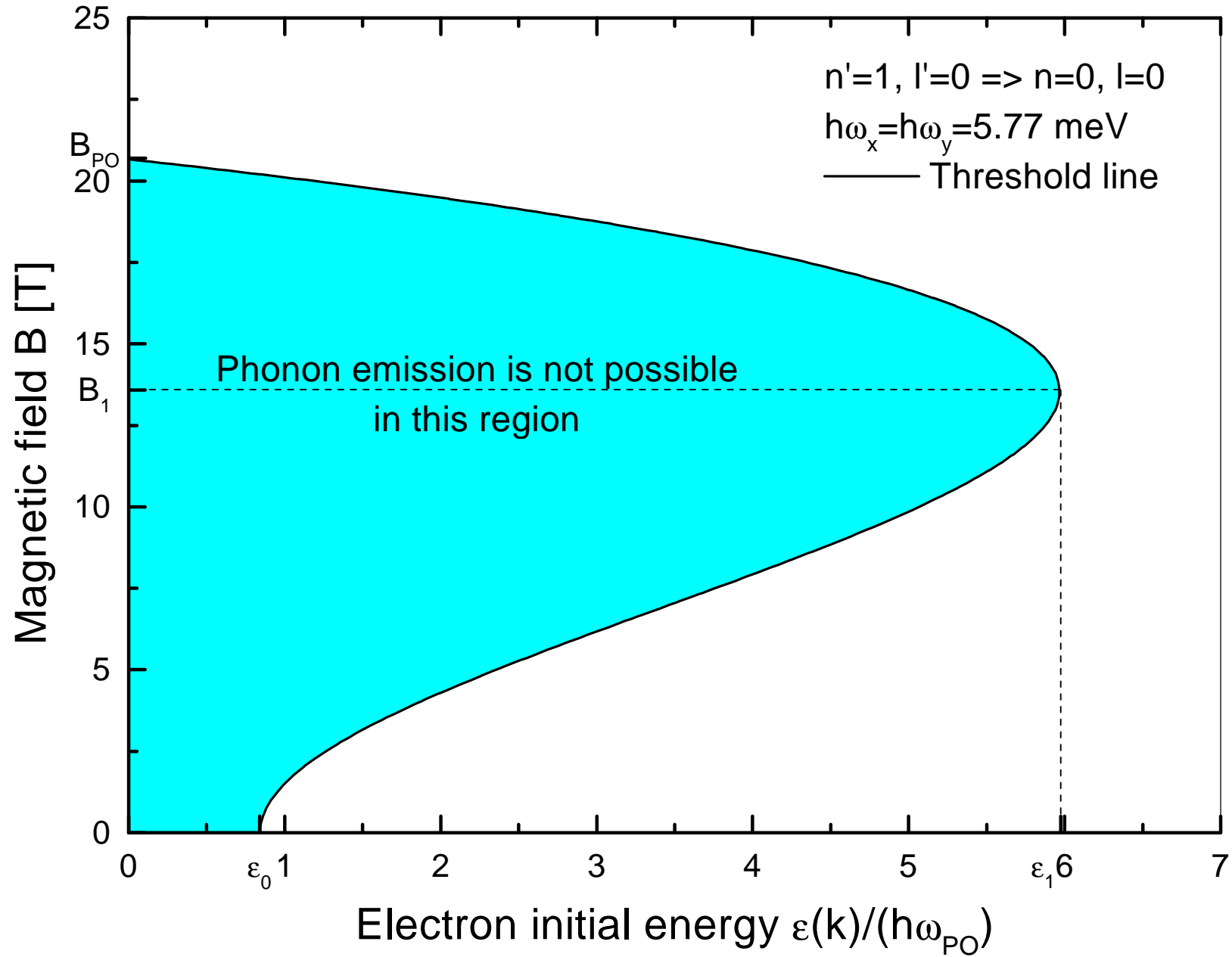
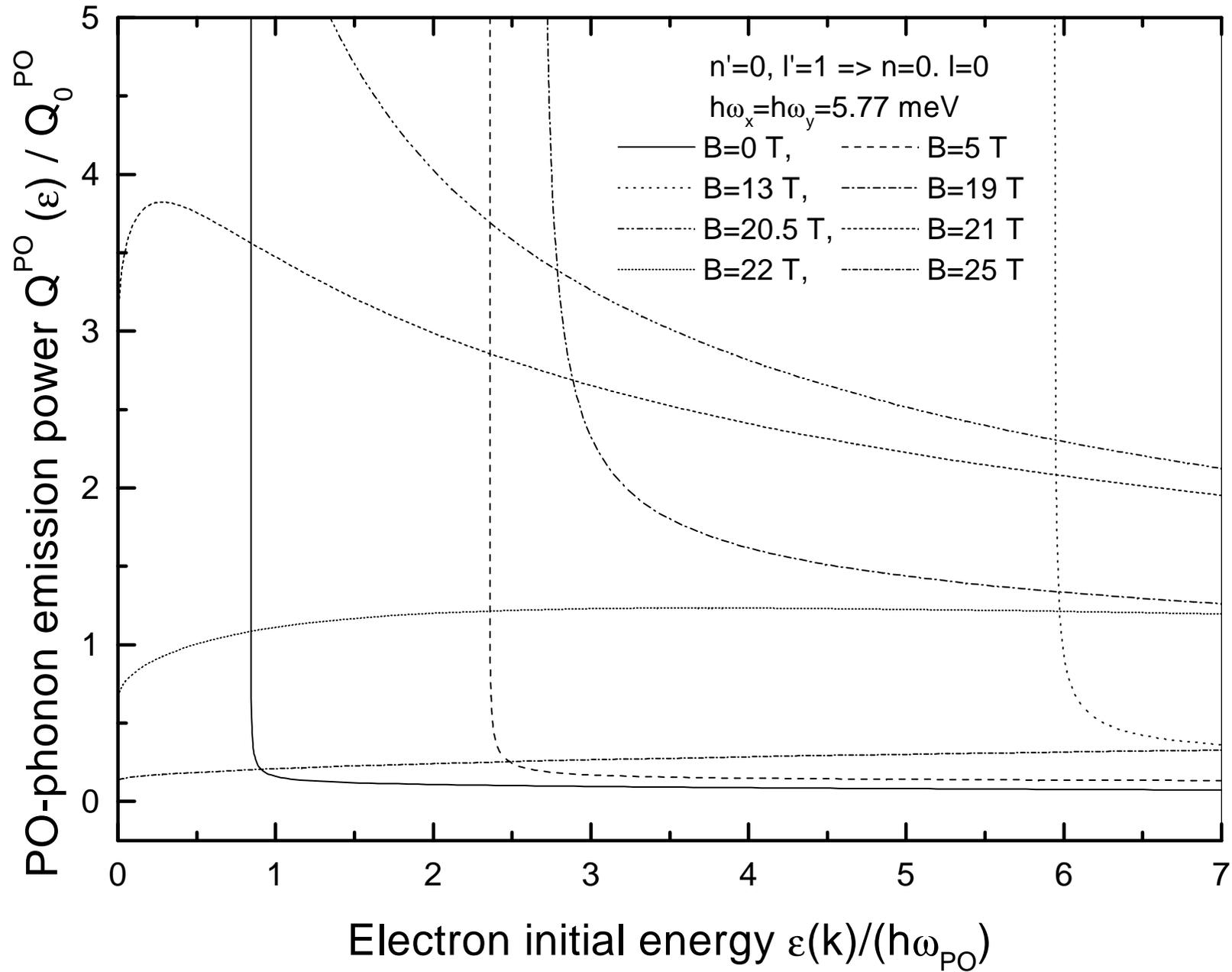
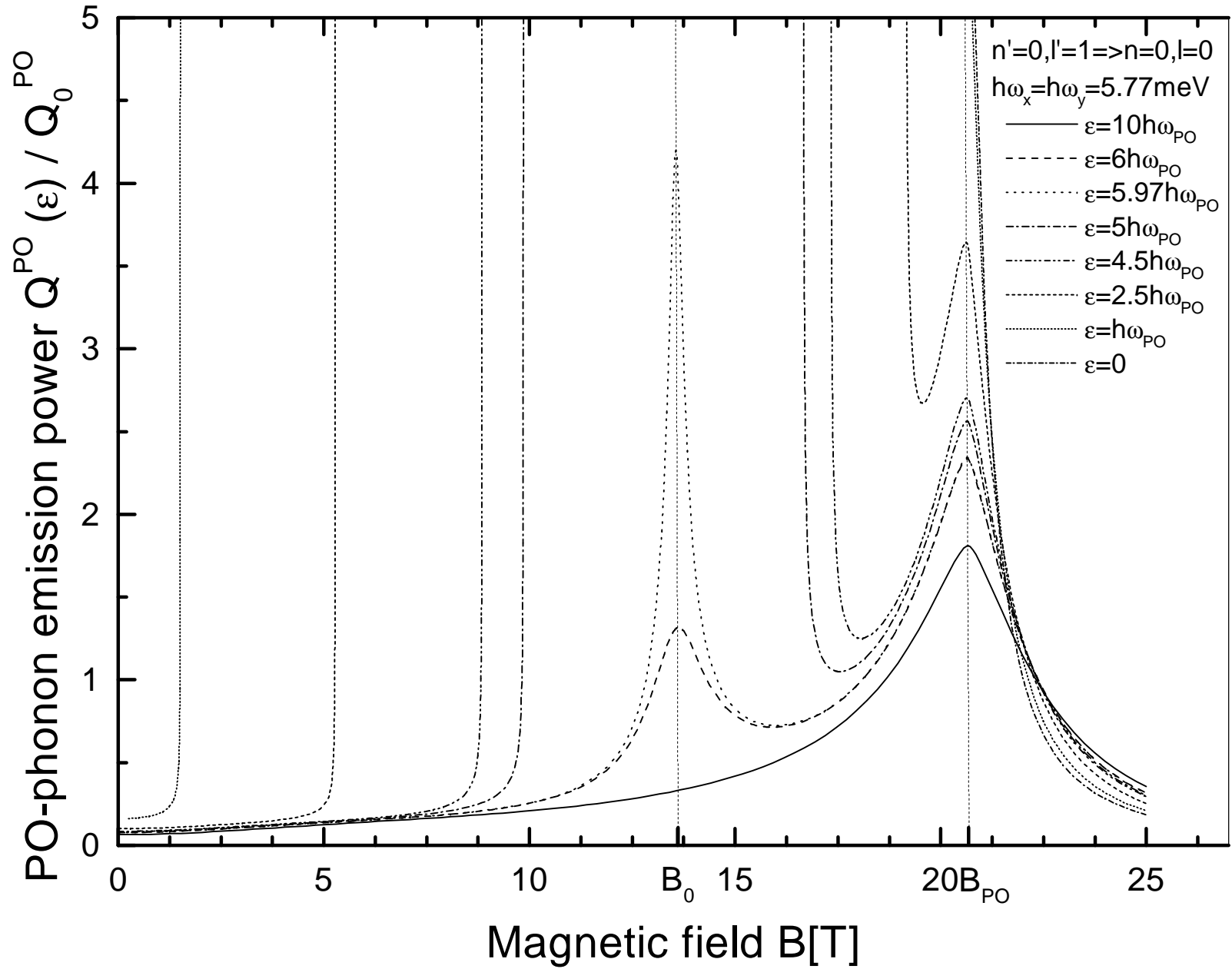
\end{document}